\documentclass[twocolumn, 
 amsmath,amssymb,
 aps,
prb,
]{revtex4-1}

\usepackage{graphicx}
\usepackage{dcolumn}
\usepackage{bm}
\usepackage{xcolor}

\emergencystretch=\maxdimen
\begin{document}


\title{Linear mapping between magnetic susceptibility and entanglement in conventional and exotic one-dimensional superfluids}

\author{D. Arisa}%
 \affiliation{Institute of Chemistry, S\~{a}o Paulo State University, 14800-090, Araraquara, S\~{a}o Paulo, Brazil}
\author{V. V. Fran\c{c}a}%
 \affiliation{Institute of Chemistry, S\~{a}o Paulo State University, 14800-090, Araraquara, S\~{a}o Paulo, Brazil}


\begin{abstract}
We investigate the mapping between magnetic susceptibility and entanglement in the metallic, insulating, conventional and exotic polarized superfluid phases of one-dimensional fermionic lattice systems as described by the Hubbard model. Motivated by recent proposals for determining and quantifying entanglement via magnetic susceptibility measurements, we numerically study the intrinsic relationship between the two quantities at zero temperature. We find signatures of the metal-insulator transition and of the BCS-BEC crossover, but the most relevant result is that for conventional and exotic superfluids the mapping between magnetic susceptibility and entanglement is surprisingly simple: {directly} proportional. This linear behavior {is found to be universal for conventional superfluids and therefore} could be exploited to quantify entanglement in current cold-atoms and condensed-matter experiments.
\end{abstract}

\maketitle

\section{\label{sec:level1}Introduction}

Entanglement, which is a theoretical concept from quantum information theory, has attracted attention from nanoscience and nanotechnology since it is considered a fundamental resource for quantum computation \cite{qc1, qc2} and quantum-enhanced metrology \cite{metro}. Entanglement has also played a central role in bridging quantum information theory to different areas, as condensed-matter, high-energy and cold-atoms physics \cite{zanardi, v2, ref1, larsson, v3, ref2, ref8, v5, kondo, ref13, ref15, ref18, v10, ref25}. By investigating entanglement properties one can probe quantum phase transitions \cite{qpt1, qpt2, malvezzi, qpt3, qpt4, v22, tiago} and characterise quantum many-body states, including exotic states of matter as Fulde-Ferrel-Larkin-Ovchnnikov superfluidity (FFLO) \cite{ff, lo, rev, v14, hulet, v17, v19}, many-body localization \cite{mb1, mb2} and topological spin liquids \cite{topo1}.

Experimentally, several protocols have been proposed to perform entanglement measurements \cite{ref13_12, ref13_13, ref13_14, ref13_15, ref13_16, greiner, tang1, tang2}, but since most of them scale with the system size exponentially, they have been restricted to few-particle systems. A possible alternative approach to determine and quantify entanglement in current experiments has been to explore intrinsic relations between entanglement and other physical quantities whose experimental measurement is well established. 

Among these quantities, we highlight the magnetic susceptibility not only because it is promptly available in cold atoms and condensed matter experiments \cite{greiner, ref13, ref25}, but also because spin and orbital fluctuations are good candidates to explain unconventional superconductivity \cite{ref10, ref10_1, ref10_3, ref10_7, ref10_9}. From a fundamental point of view, it seems reasonable to expect strong connections between entanglement and magnetic susceptibility, because the latter is also close connected to another concept from quantum information theory, the fidelity. Fidelity is a measure of the similarity between two quantum states with respect to a driving parameter and its most relevant term $-$ the fidelity susceptibility  \cite{ref1, ref2, ref8, ref15, ref11} $-$ is essentially the magnetic susceptibility for thermal states when the driving parameter is an external magnetic field or an internal magnetization.

Although there are several works connecting entanglement to quantum phase transitions and magnetic susceptibility or more general fidelity susceptibility to quantum phase transitions, just a few works directly relate entanglement to magnetic susceptibility in multiband topological insulators \cite{ref8}, Ising models \cite{ref13} and spin chains \cite{ref15, ref25, ref18}. For the fermionic Hubbard model, in particular, reports associating entanglement to magnetic susceptibility are restricted to half-filled systems with strong repulsive interactions \cite{v1}, regime that actually make the Hubbard model equivalent to weakly interacting spin chains \cite{v6}. 

Here we investigate the intrinsic relationship between entanglement and magnetic susceptibility in fermionic systems in the metallic, insulating, conventional and exotic (FFLO) superfluid phases. Our analysis $-$ within the single-band one-dimensional Hubbard model at zero temperature $-$ reveals that the mapping between entanglement and magnetic susceptibility in conventional and exotic superfluids is as simple as it could be: {\it {directly} proportional}. To our knowledge this linearity has not been reported in the literature. By analysing its features and peculiarities, we explain the linear mapping between entanglement and magnetic susceptibility, demonstrating that it is not an artefact or a coincidence. {We thus determine and comprise in Eq.(7) the universality of this linear behavior for conventional superfluids, allowing thus the quantification of} entanglement in cold-atoms and condensed-matter experiments.

\begin{figure*}[!t]
 \centering

\hspace{-0.3cm}\includegraphics[scale=0.3]{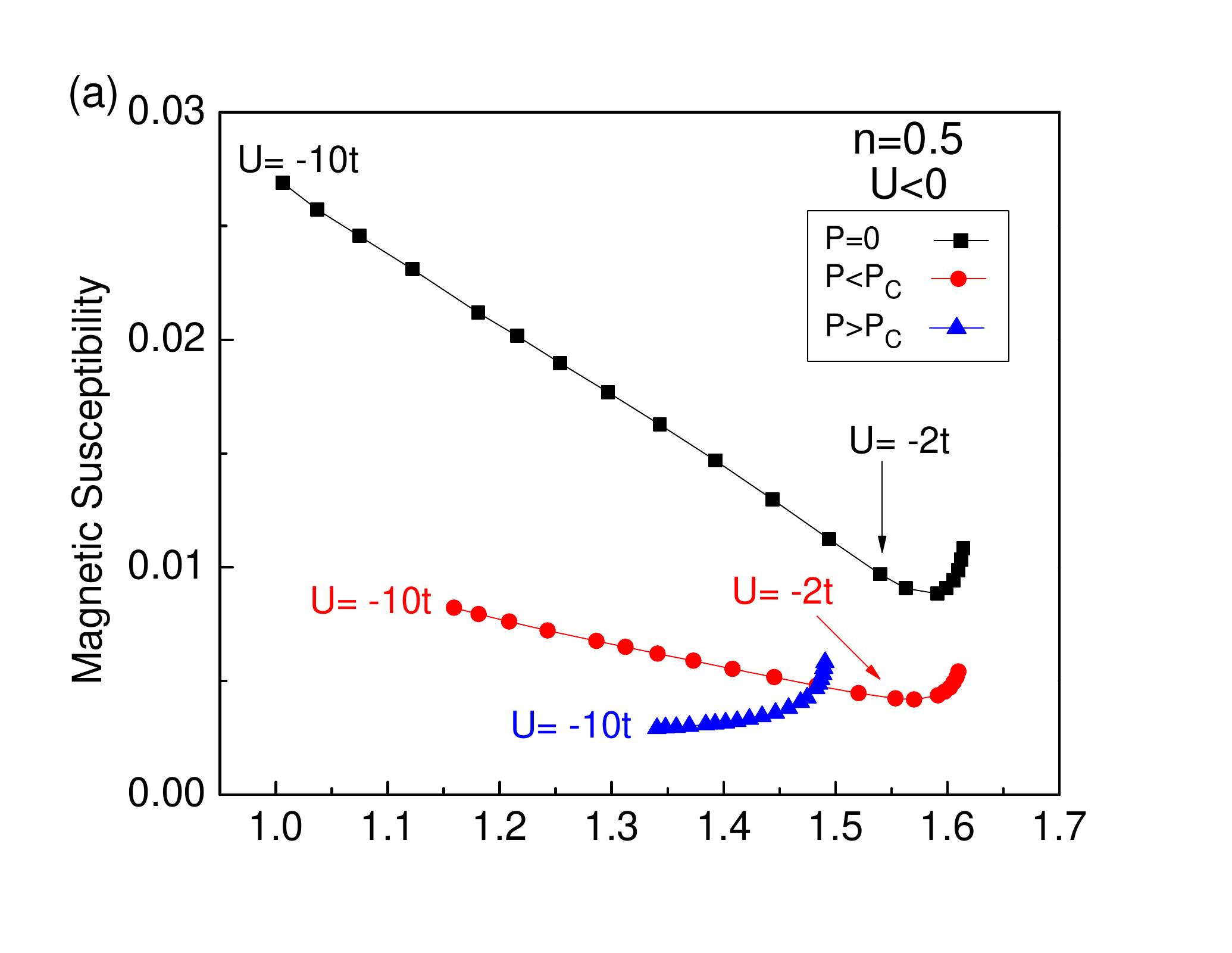}\hspace{-1cm}\includegraphics[scale=0.3]{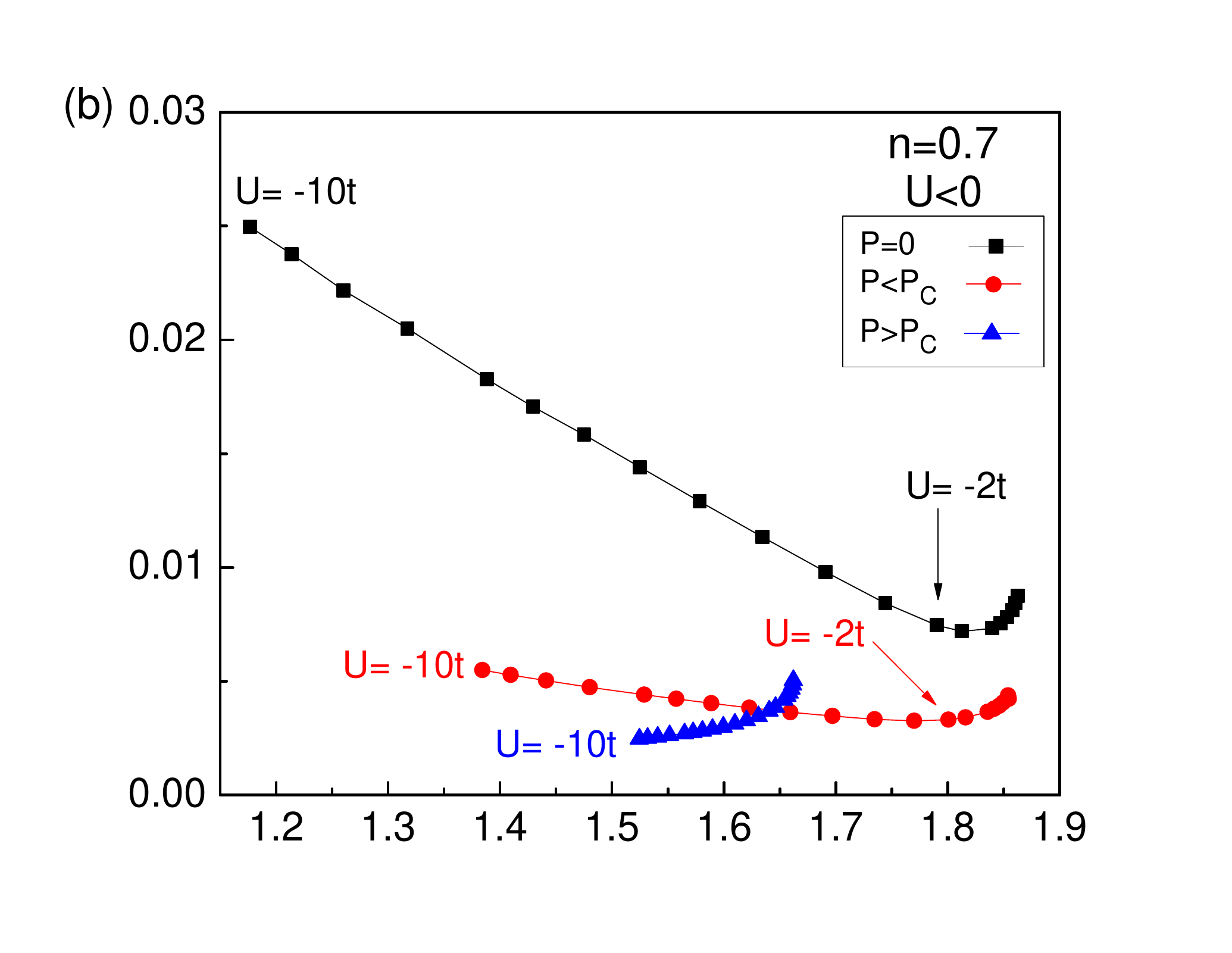}\hspace{-1cm}\includegraphics[scale=0.3]{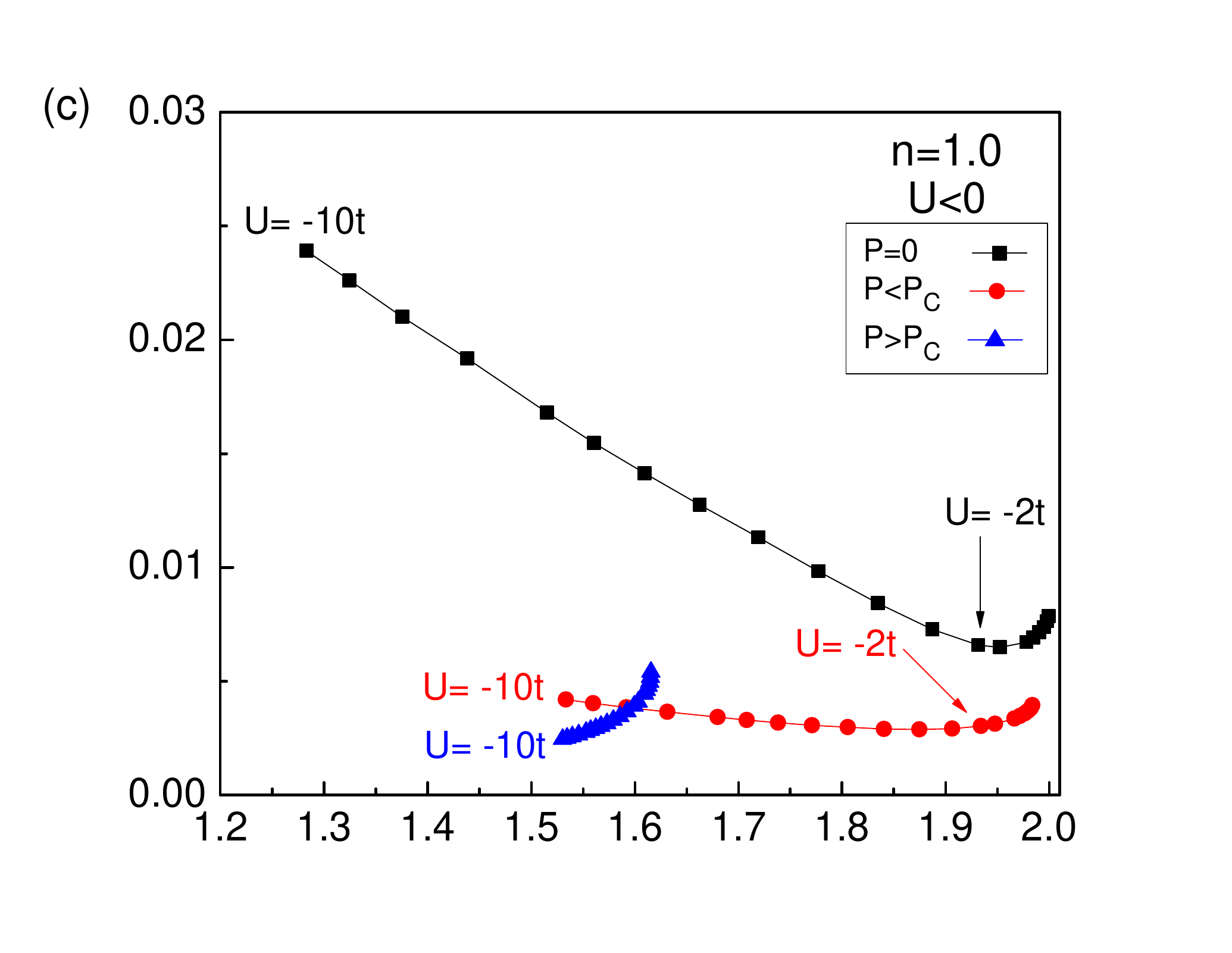}\vspace{-0.8cm}

\hspace{-0.3cm}\includegraphics[scale=0.3]{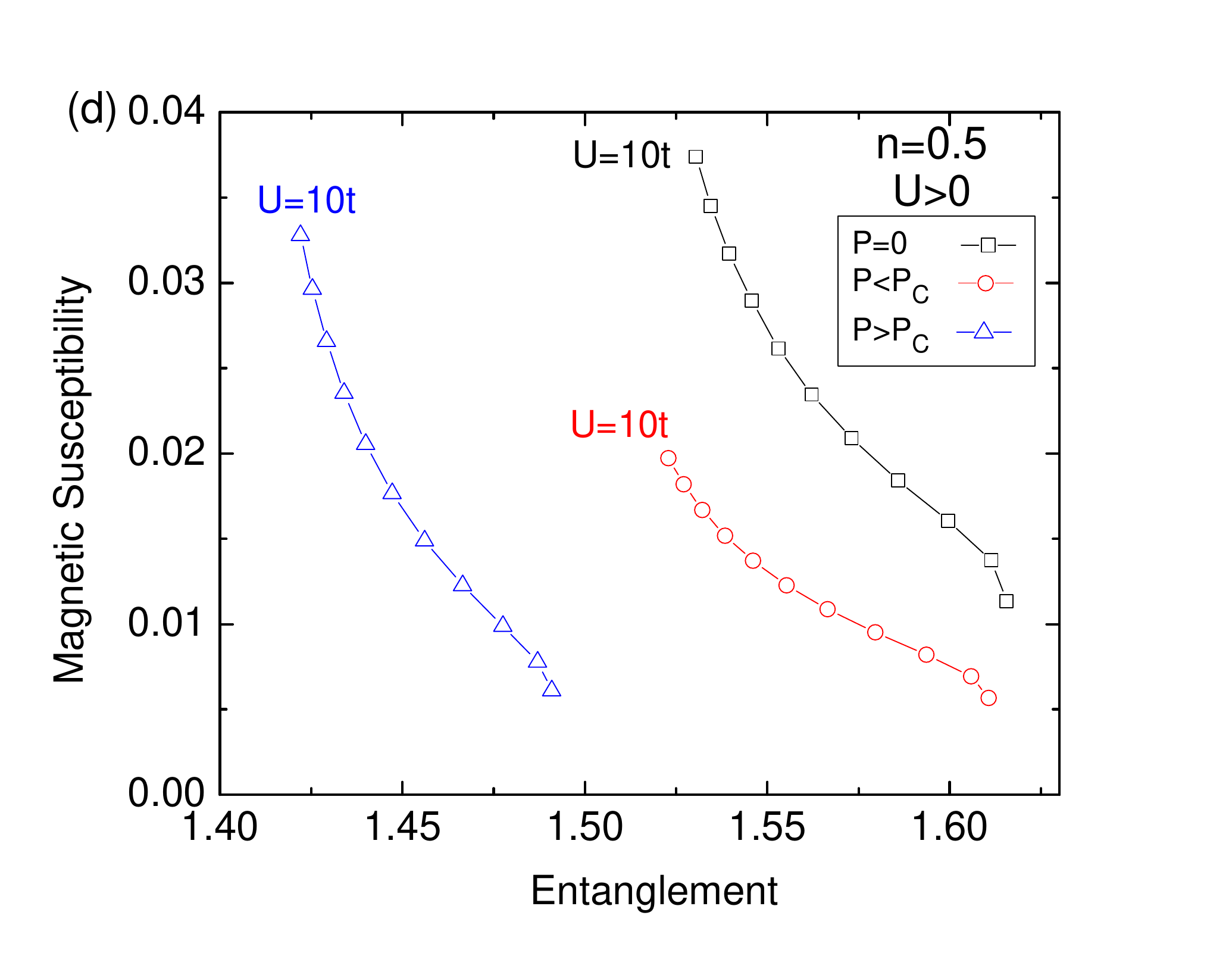}\hspace{-1cm}\includegraphics[scale=0.3]{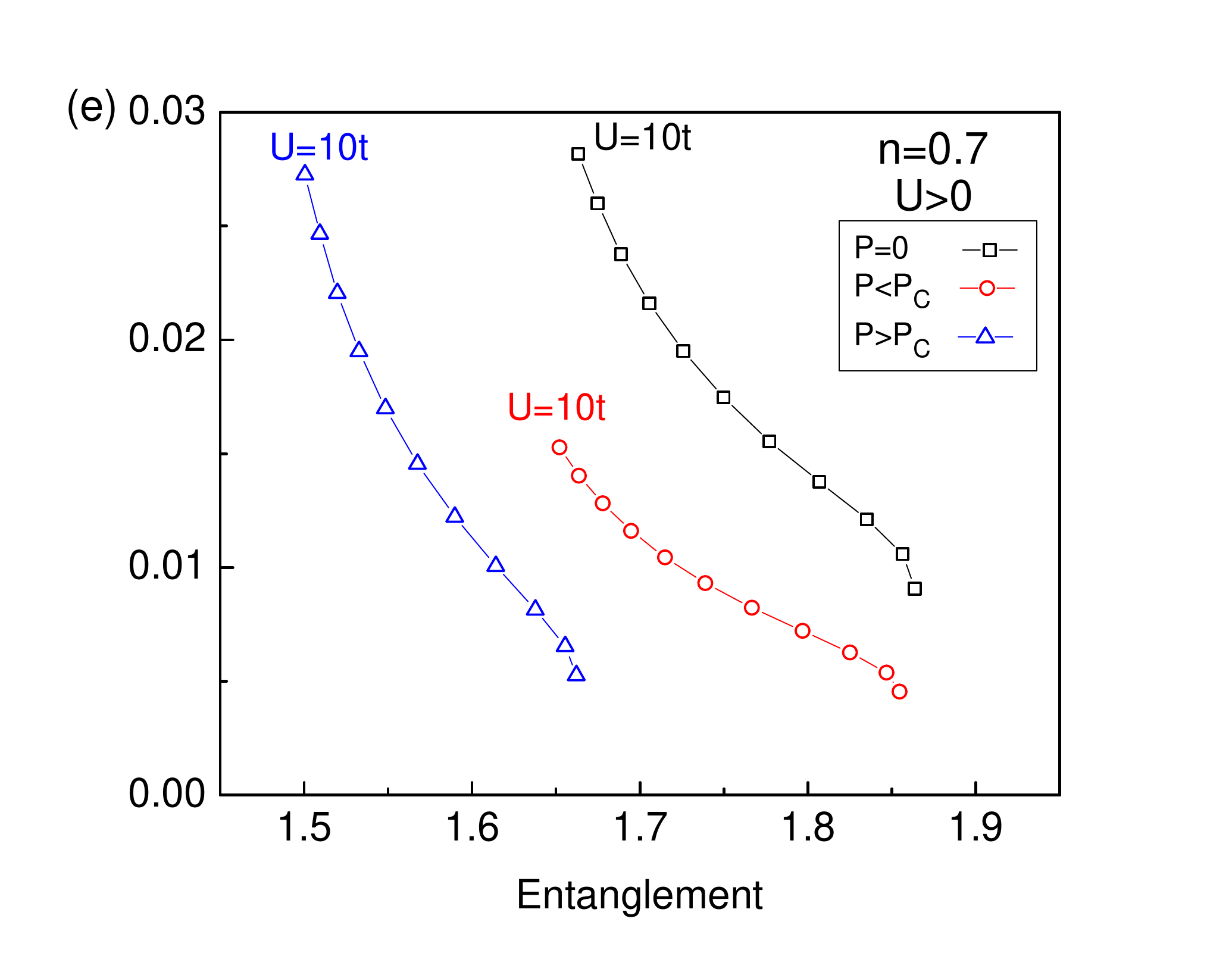}\hspace{-1cm}\includegraphics[scale=0.3]{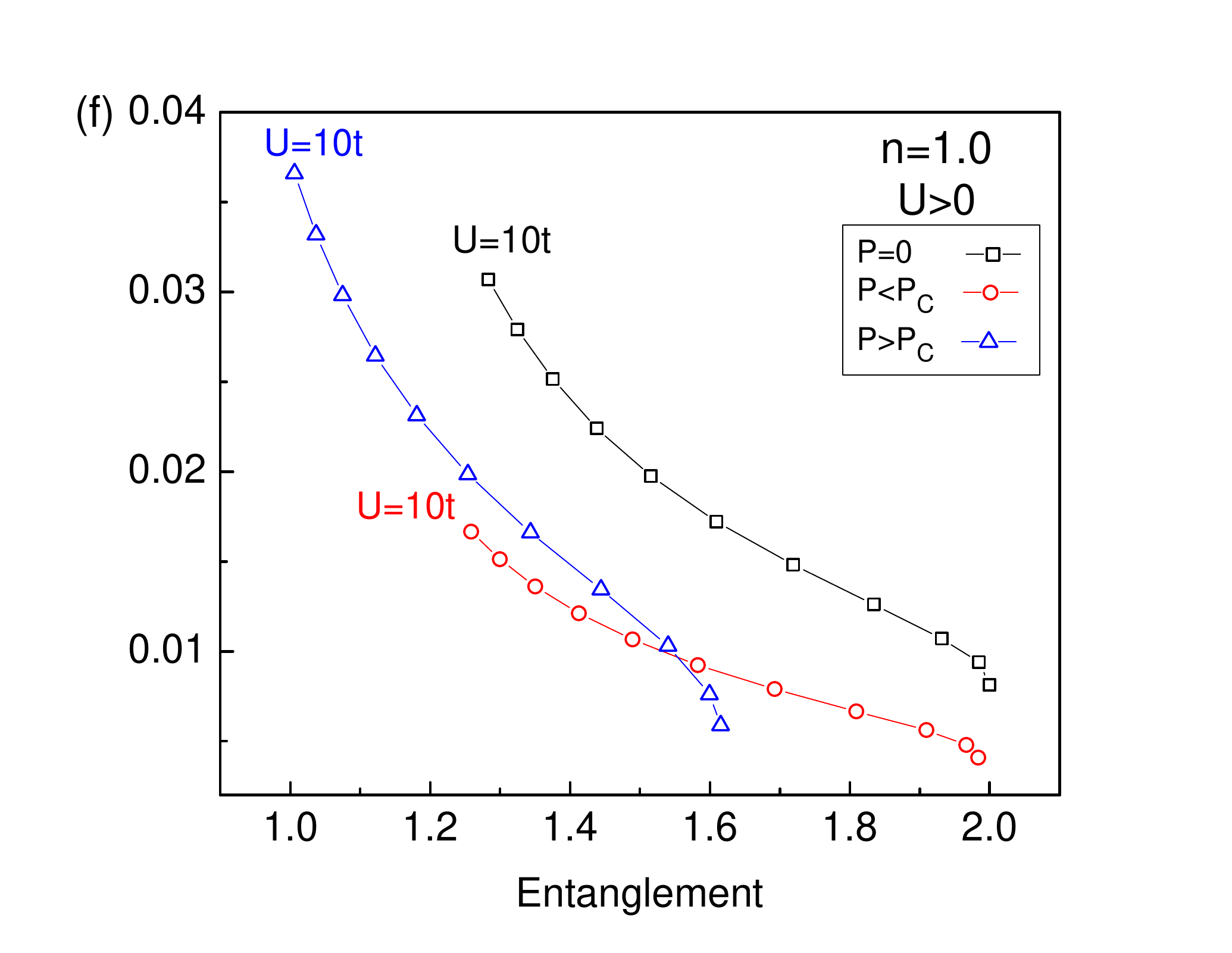}\vspace{-0.5cm}

           \caption{Magnetic susceptibility as a function of entanglement for attractive ({upper panels}) and repulsive ({bottom panels}) interactions, for $P=0$, $P<P_C$ ($P=0.1$) and $P>P_C$ ($P=0.5$), for $n=0.5$, $n=0.7$ and $n=1.0$. We indicate the initial and final corresponding interaction, which ranges from $U=-10t$ to $U=10t$, {and highlight the point} $U=-2t$ for $0\leq P<P_C$ {which delimitates} the linear behavior found for $U\lesssim -2t$. }\label{fig1}
   \end{figure*}

\section{\label{sec:level2}Theoretical Model and Computational Methods}

We consider one-dimensional nanostructures at zero temperature as described by the single-band fermionic Hubbard model,
\begin{eqnarray}
H=-t\sum_{<ij>\sigma}\hat c_{i\sigma}^{\dagger}\hat c_{j\sigma}
 +U\sum_{i}\hat{n}_{i\uparrow}\hat{n}_{i\downarrow},
\label{eqn:HubbardHamiltonian}
\end{eqnarray}
where $U$ is the on-site interaction, $t$ the hopping parameter between neighbour sites $<ij>$, $\hat n_{i,\sigma}=\hat c_{i,\sigma}^{\dagger}\hat c_{i,\sigma}$ the density operator and $\hat c_{i,\sigma}^{\dagger}$ ($\hat c_{i,\sigma}$) the creation (annihilation) operator of fermionic particles with 
$z$-spin component $\sigma=\uparrow,\downarrow$ at site $i$. The filling factor or average density is given by $n=N/L$, while the magnetization by $m=n_\uparrow - n_\downarrow$, where $N=N_\uparrow+N_\downarrow$ is the total number of particles and $L$ the chain size. Throughout this paper we consider $t = 1$, fixed total number $N$ of particles, $L=80$ and open boundary conditions. 

 \begin{figure*}[t]
 \centering
\hspace{-0.3cm}\includegraphics[scale=0.3]{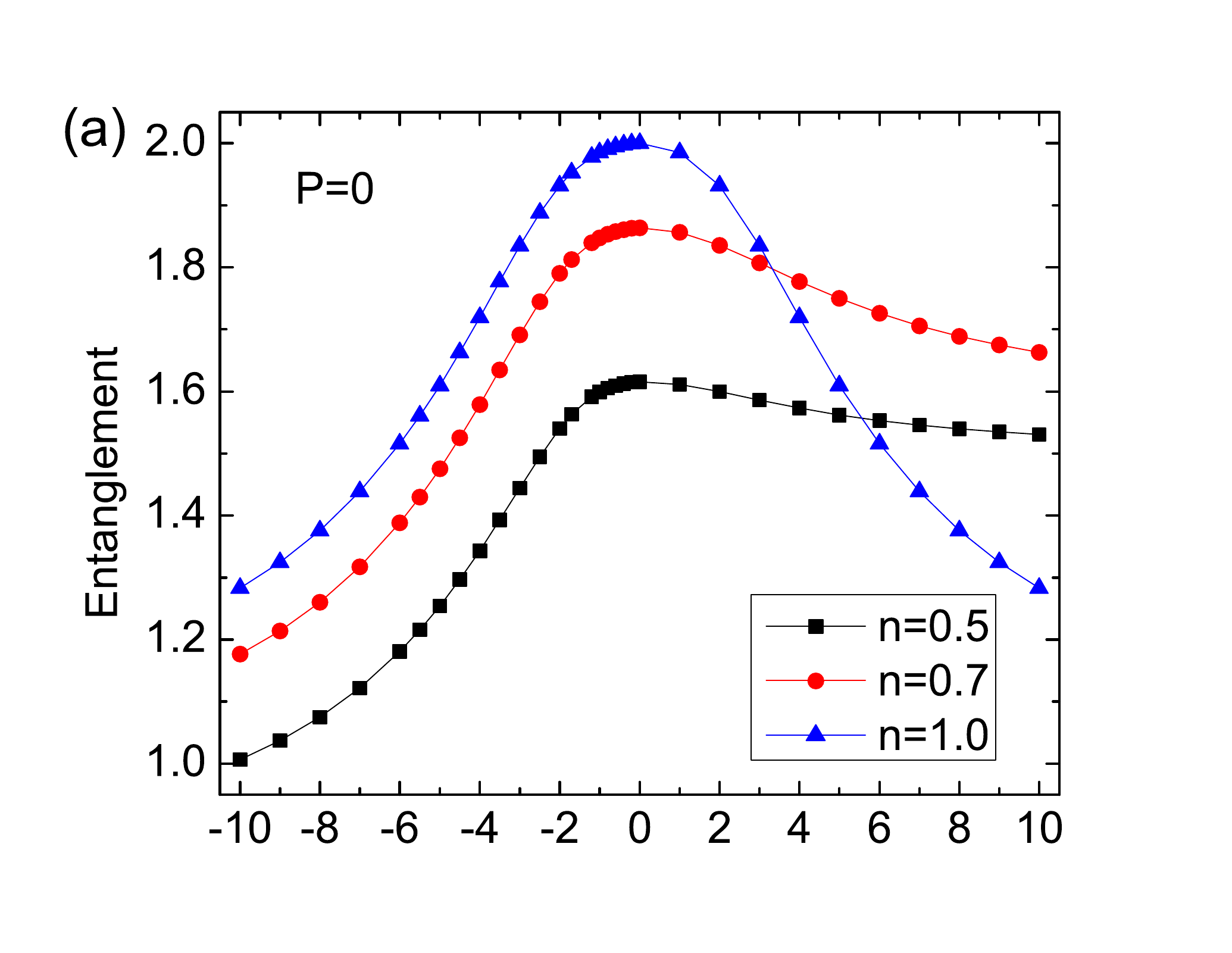}\hspace{-1cm}\includegraphics[scale=0.3]{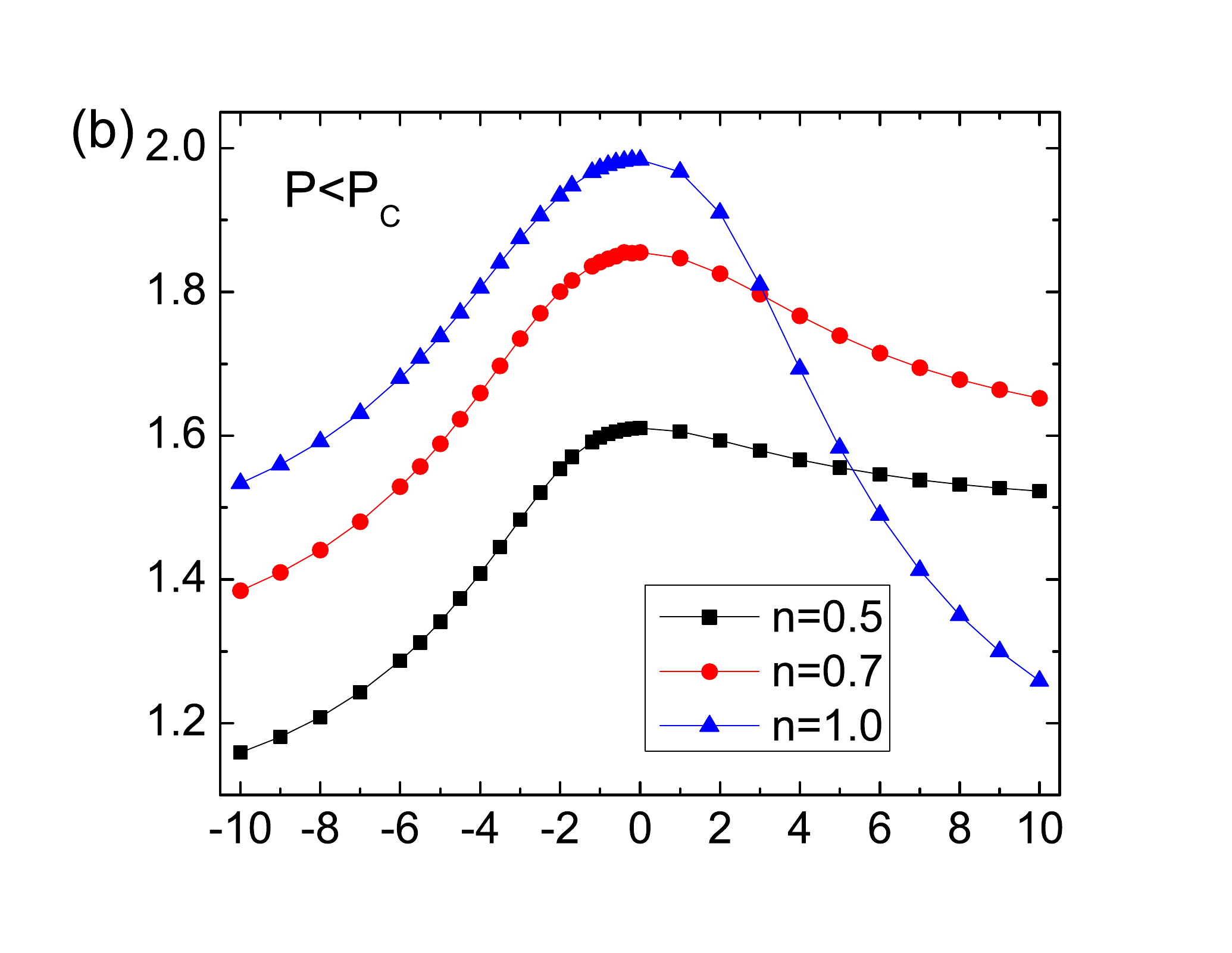}\hspace{-1cm}\includegraphics[scale=0.3]{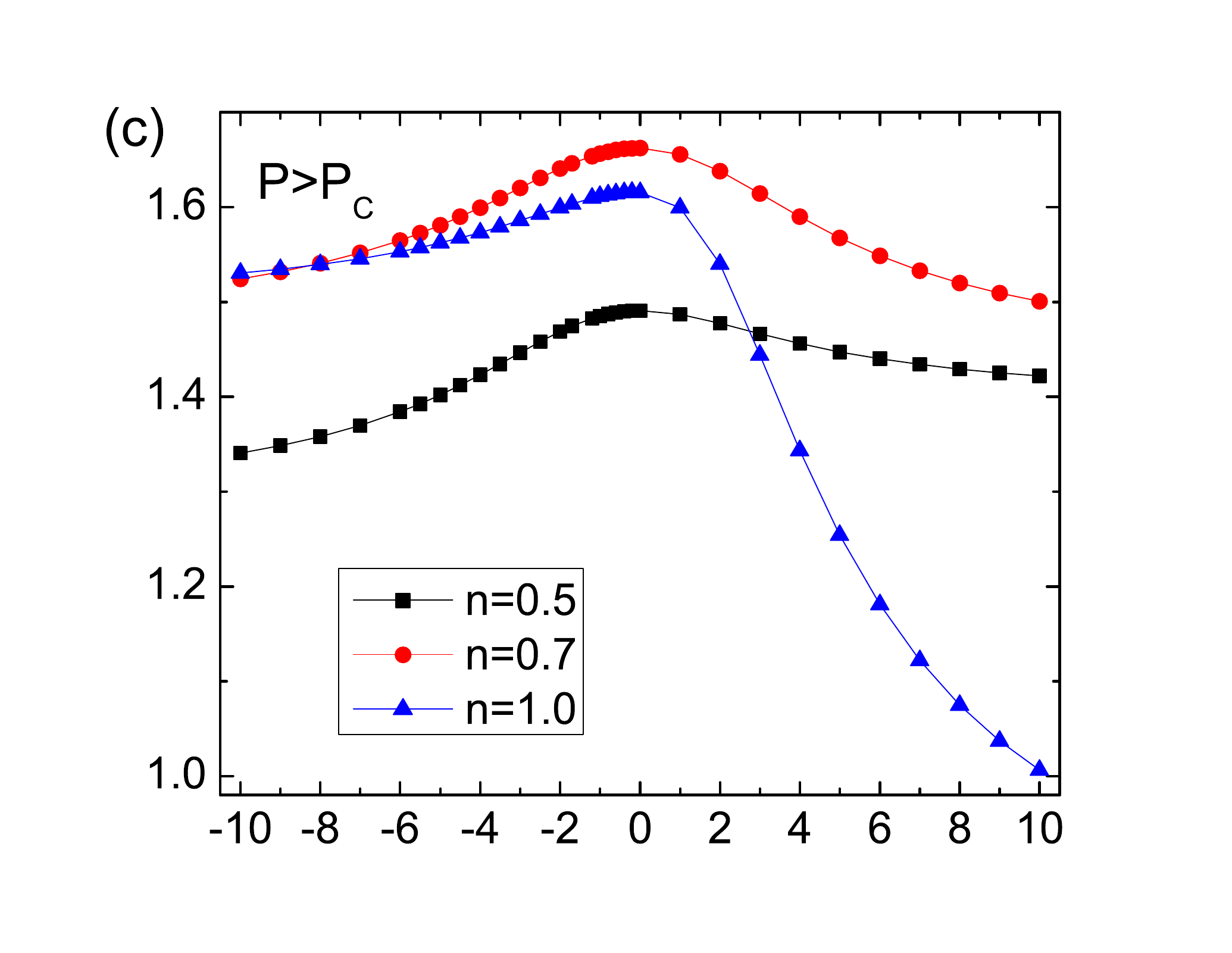}\vspace{-0.8cm}

\hspace{-0.3cm}\includegraphics[scale=0.3]{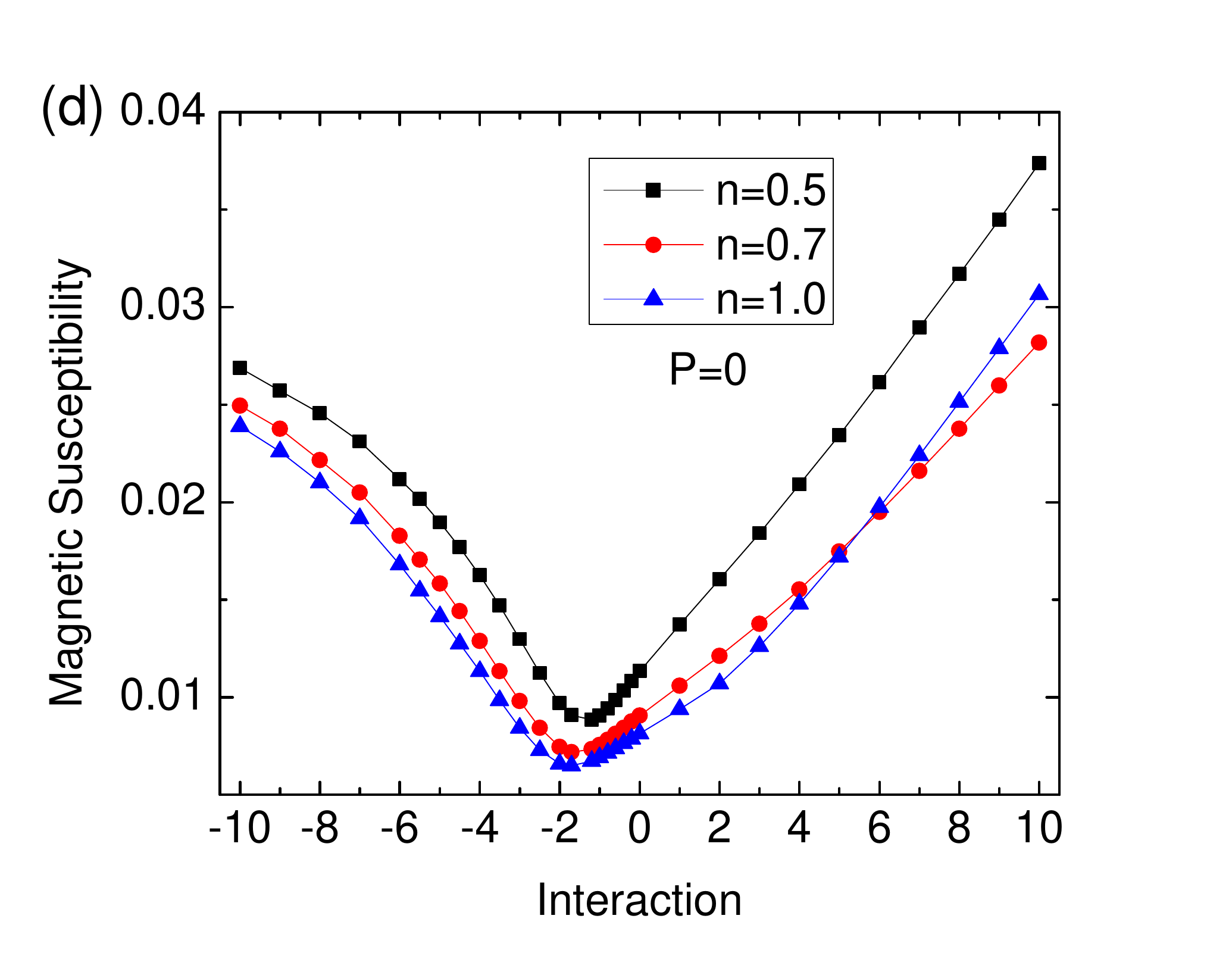}\hspace{-1cm}\includegraphics[scale=0.3]{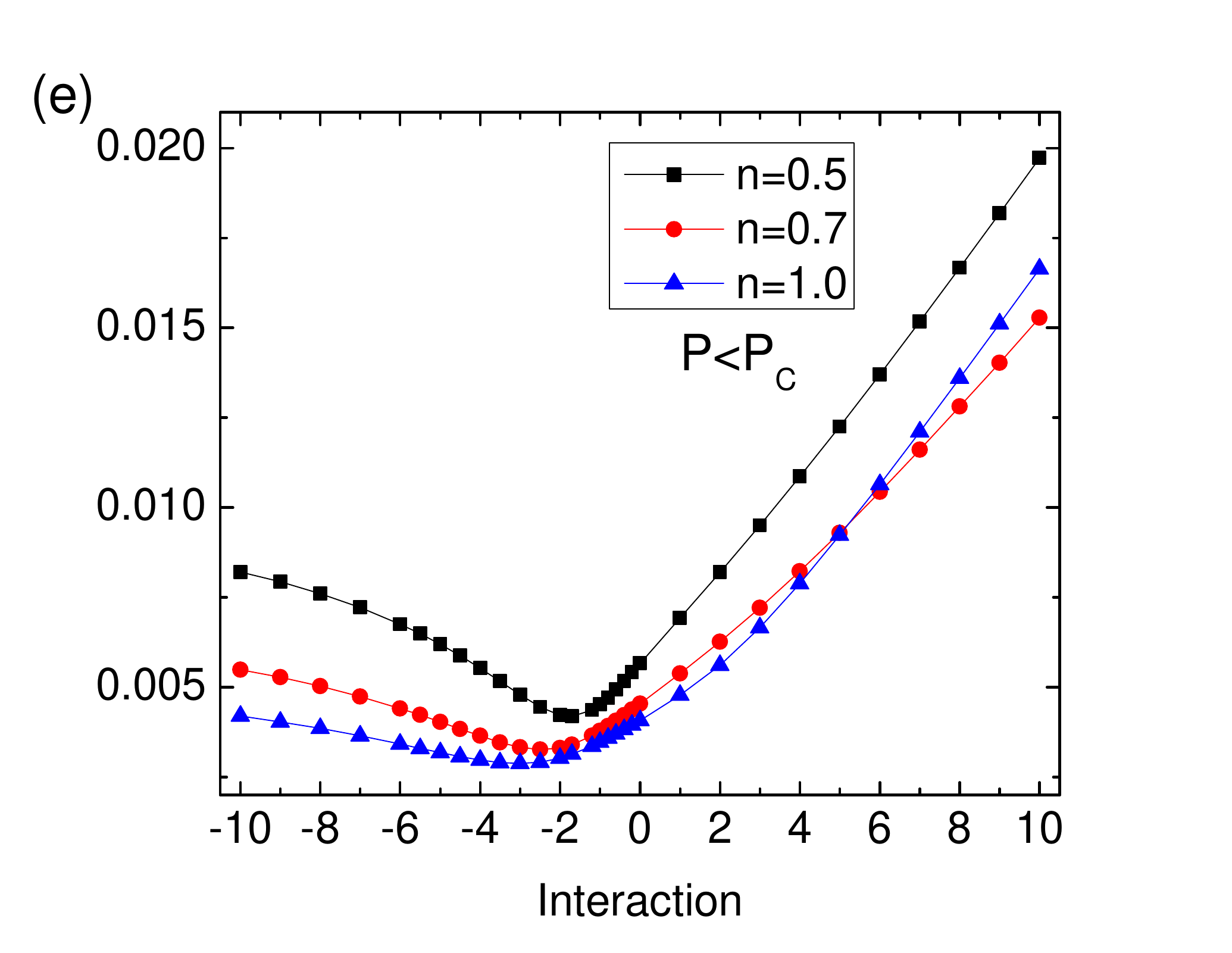}\hspace{-1cm}\includegraphics[scale=0.3]{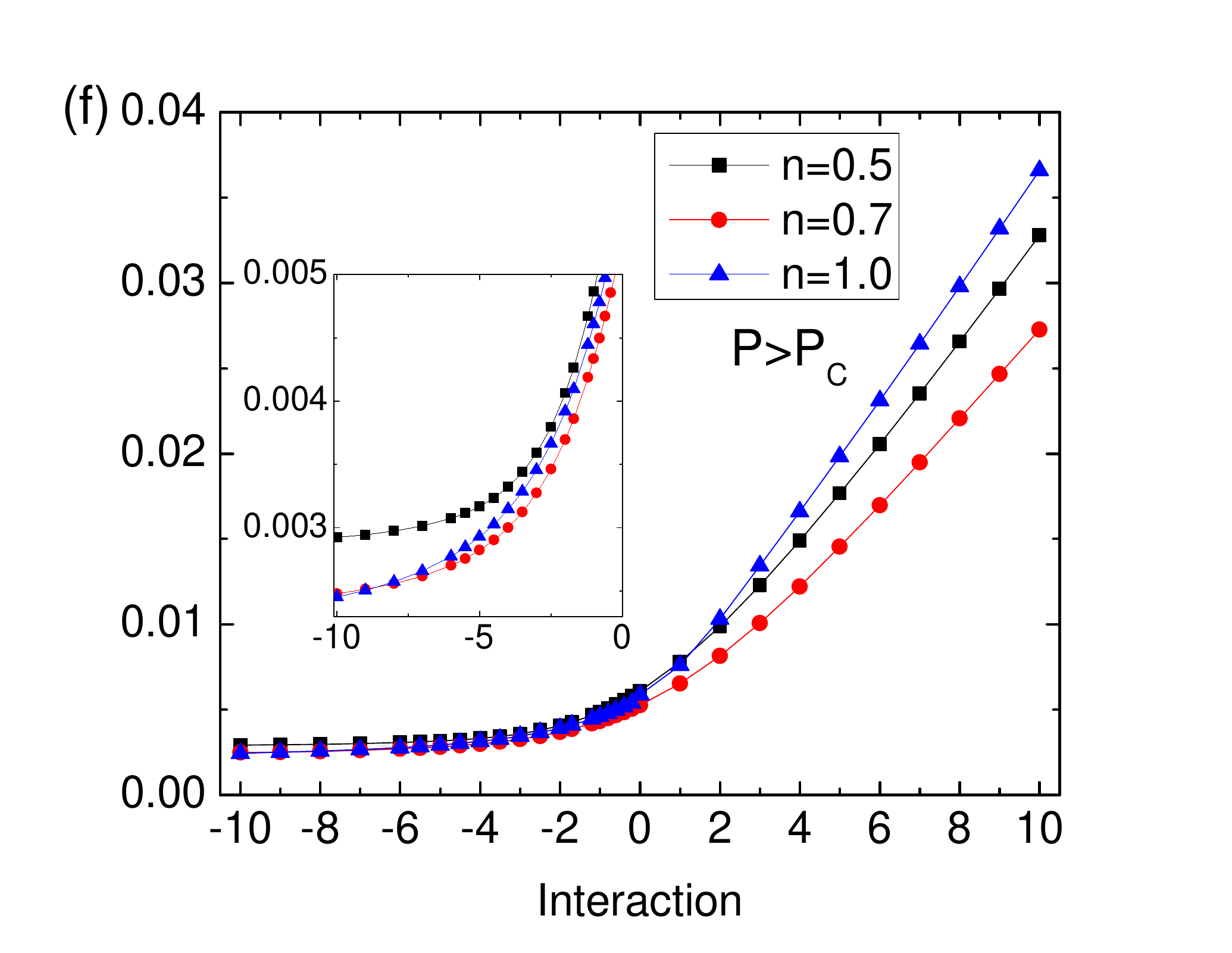}\vspace{-0.5cm}
       \caption{Entanglement (upper panels) and magnetic susceptibility (lower panels) as a function of interaction: (a) and (d) for $P=0$, (b) and (e) for $P<P_C$ ($P=0.1$) and (c) and (f) for $P>P_C$ ($P=0.5$). Inset is just zoom in.}

     \end{figure*}

We determine the single-site entanglement in such chains at the ground state via the von Neumann entropy,
\begin{eqnarray}
S_i&=&-w_{\uparrow,i} \log_2 w_{\uparrow,i}-w_{\downarrow,i} \log_2 w_{\downarrow,i}\nonumber\\
&&-w_{2,i} \log_2 w_{2,i}-w_{0,i} \log_2 w_{0,i},\label{eq.linear}
\end{eqnarray}
a well-defined entanglement measure for bipartite pure systems \cite{qpt2}, which quantifies the entanglement between site $i$ and the remaining $L-1$ sites. Here $w_{2,i}$ is the double occupancy probability at site $i$,
\begin{equation}
w_{2,i}=\frac{\partial e_0(n,m,U)}{\partial U},
\end{equation}
$w_{\uparrow, i}=n/2+m/2-w_{2,i}$ and $w_{\downarrow, i}=n/2-m/2-w_{2,i}$ are the single-particle or unpaired probabilities, and \hspace{0.5cm}$w_{0,i}=1-w_{\uparrow,i}-w_{\downarrow,i}-w_{2,i}$ is the zero-occupation probability. Here $e_0(n,m,U)=E_0(n,m,U)/L$ is the per-site ground-state energy. In order to avoid site-dependent quantities, our results for entanglement and doubly-occupied probability are averaged over the chain sites: $S\equiv \sum_i S_i/L$ and $w_2\equiv \sum_i w_{2,i}/L$.

\begin{table}[!b]
\centering
\begin{tabular}{|ll|ll|}
\hline
& ${\bf U>0}$ &  & ${\bf U<0}$ \\
 $n\neq1$ &  metal &   P=0 &  conventional superfluid\\
 $ n=1$ &  insulator  &  $P<P_C$ &  exotic superfluid\\
&  &  $P>P_C$ &  normal non-superfluid\\
\hline
\end{tabular}
\normalsize
\caption{Summary of the main phases of the Hubbard model for a given interaction $U$, density $n$ and polarization $P$.}
\end{table}

The magnetic susceptibility $\chi$ is numerically obtained through the second derivative of the total energy $E_0(n,m,U)$ with respect to the magnetization at fixed density and interaction,
\begin{equation}
\chi = \left[\frac{\partial^2 E_0(n,m,U)}{\partial m^2}\right]_{n,U}^{-1}.
\end{equation}
Ground-state energies and occupation probabilities are obtained with density-matrix renormalization group (DMRG) \cite{dmrg} techniques, therefore our results are numerically exact. Alternative approaches would be for example {\it i)} to obtain approximate results by performing density-functional theory calculations \cite{review, v16} or  {\it ii)} numerically solve the Lieb-Wu equations \cite{lw} for infinite chains. However for the development of experimental nanophysics and nanotechnology it is crucial to obtain theoretical results of finite systems. Additionally, as $n, m$ are in the Lieb-Wu integrals indirect parameters, it would be necessary a huge amount of data to generate the derivative in Eq. (4) at fixed $n,U$. 

The polarization due to the imbalance between spin populations is given by 
$P=(N_\uparrow-N_\downarrow)/N$, which is related to the magnetization by {$m=nP$}. For attractive interactions, $U<0$, the system presents conventional BCS \cite{bcs} superfluidity for $P=0$, exotic FFLO superfluidity for $P<P_C$ and a normal non-superfluid phase for $P>P_C$, where $P_C$ is the critical polarization delimitating the FFLO to the normal phase. One can obtain $P_C$ via the equality \cite{v14}
\begin{equation}
P_C(n,U)= \frac{4w_2(n,P_C,U)}{n}-1.
\end{equation}
Notice that although $P_C$ depends on $n$ and $U$, it has a universal upper bound \cite{v14} given by  $P_C^{max}=1/3$. 

For repulsive interactions, $U>0$, the system is either a metal for $n\neq 1$ or an insulator for $n=1$: at $U=0$ and half filling the system undergoes the Mott transition from an ideal conductor to an insulating phase. Table 1 summarizes the different physical phases we will consider within the Hubbard chains.
\begin{figure*}[!t]
 \centering

\hspace{-0.3cm}\includegraphics[scale=0.3]{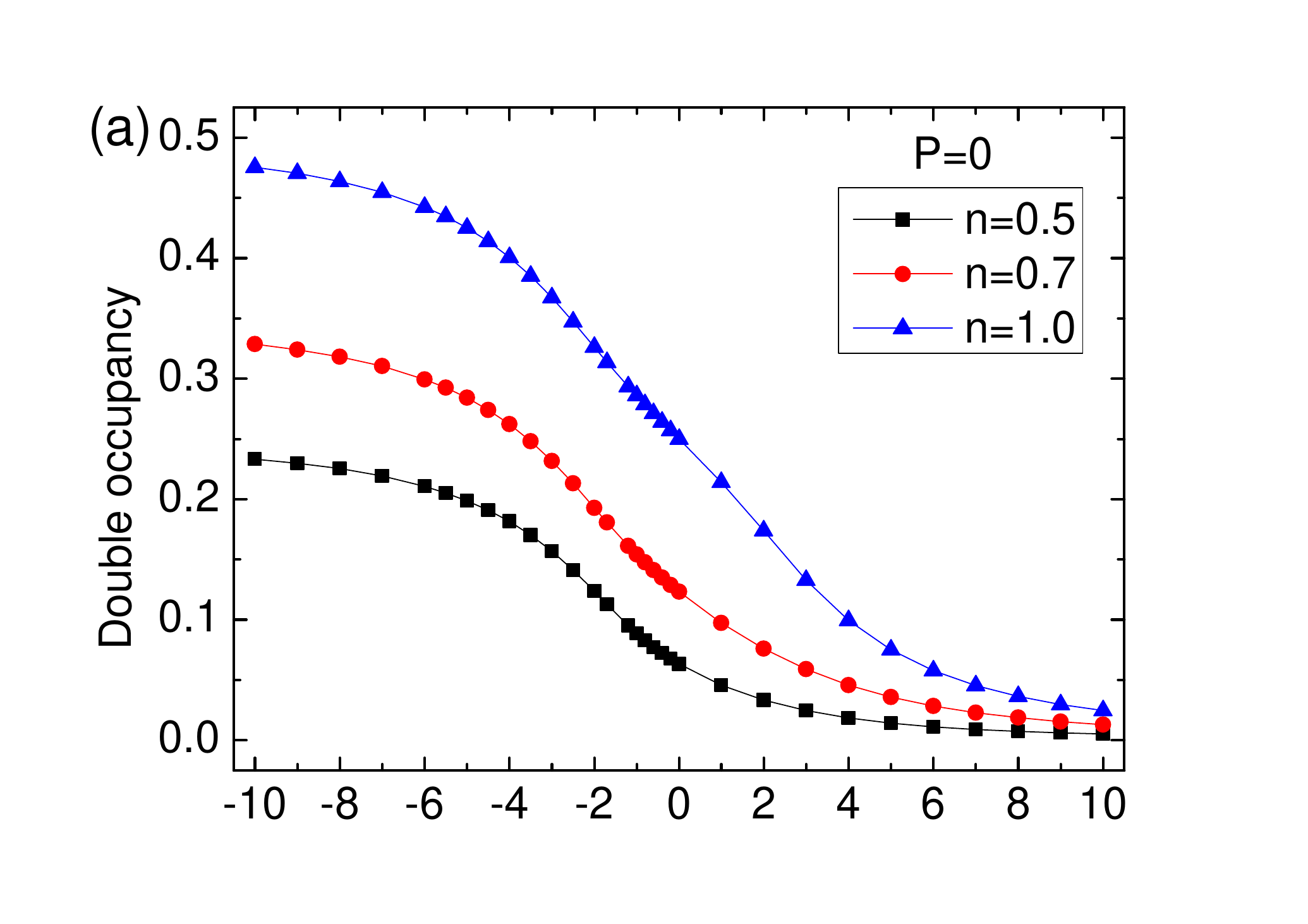}\hspace{-1cm}\includegraphics[scale=0.3]{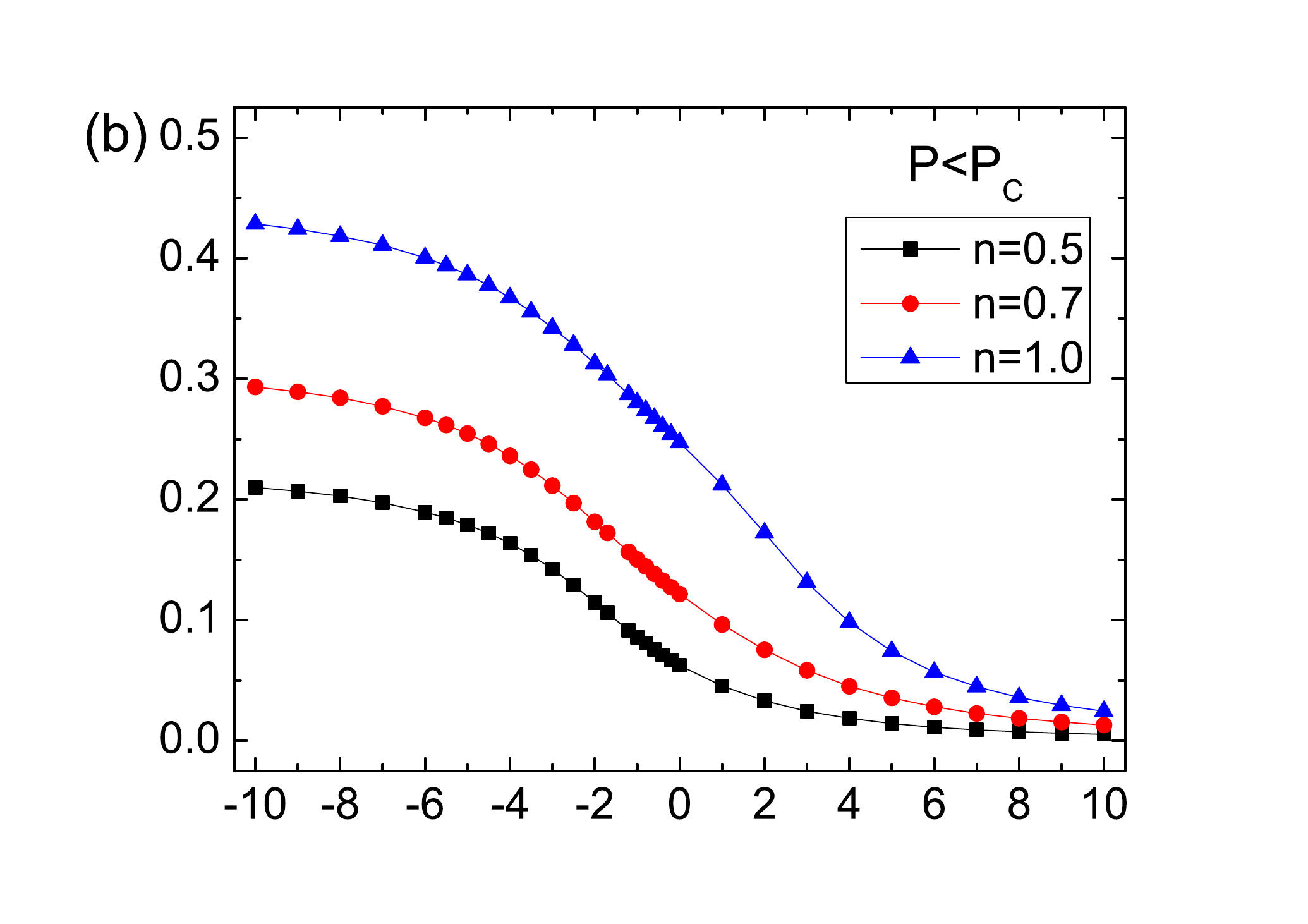}\hspace{-1cm}\includegraphics[scale=0.3]{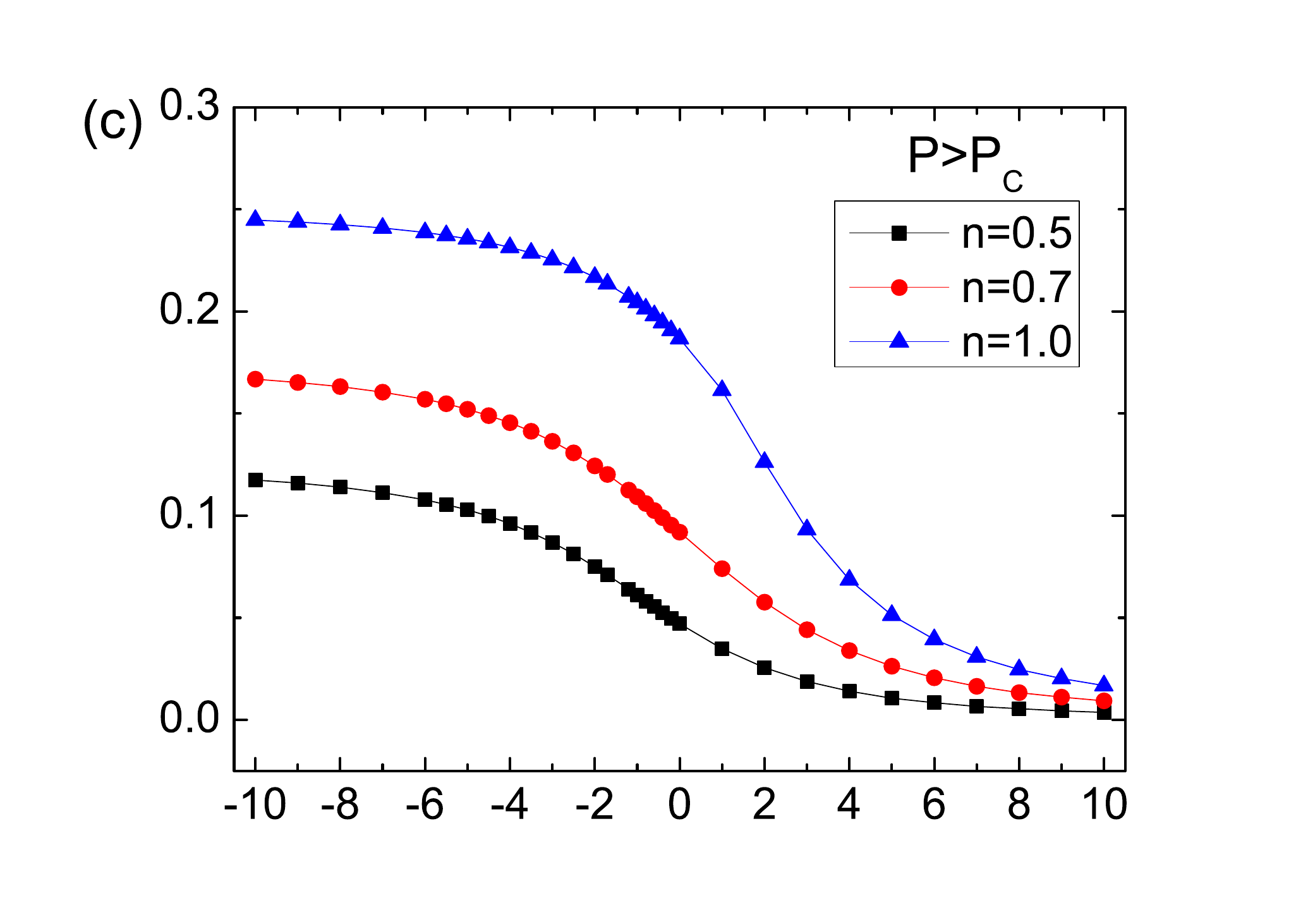}\vspace{-0.8cm}

\hspace{-0.3cm}\includegraphics[scale=0.3]{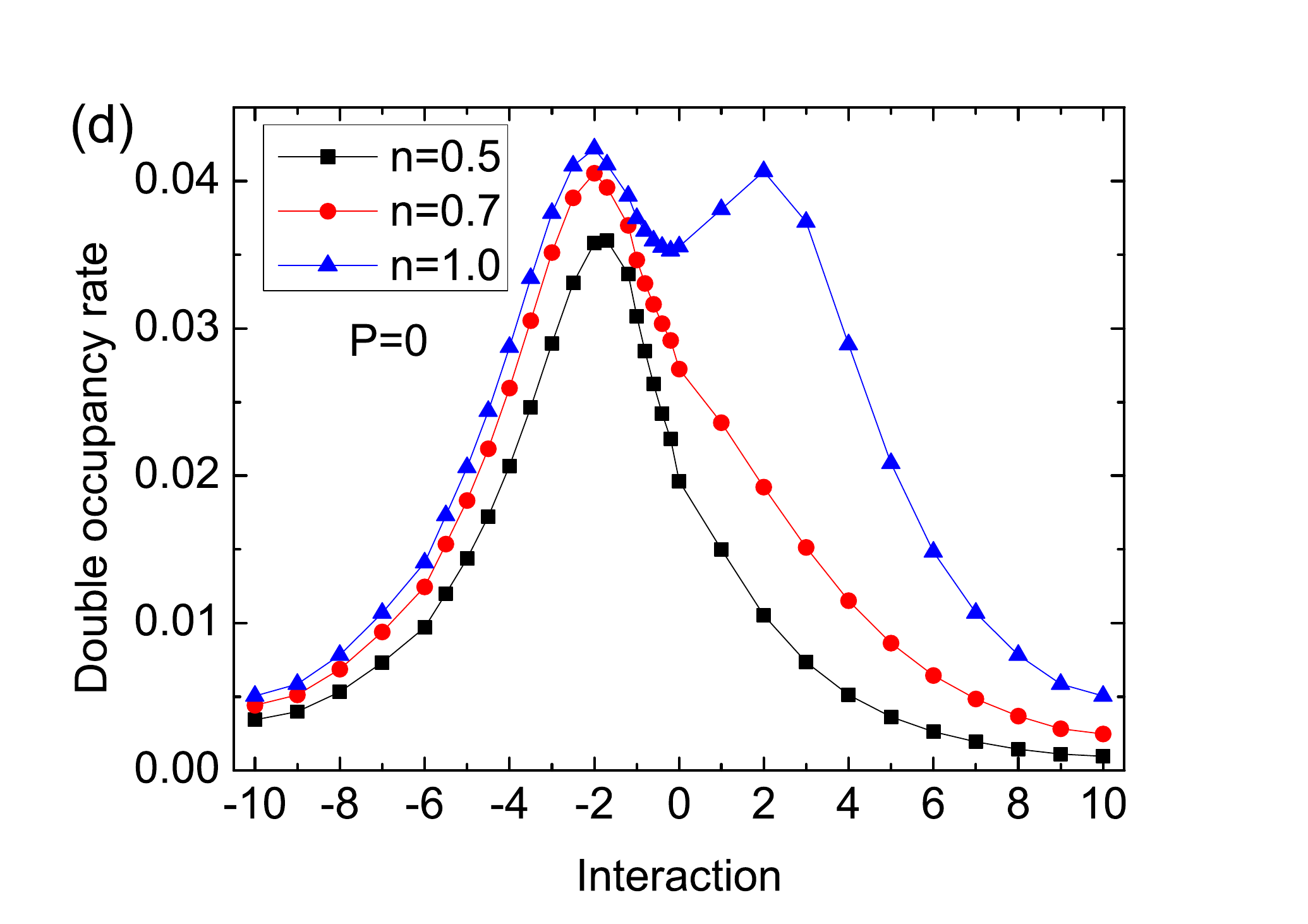}\hspace{-1cm}\includegraphics[scale=0.3]{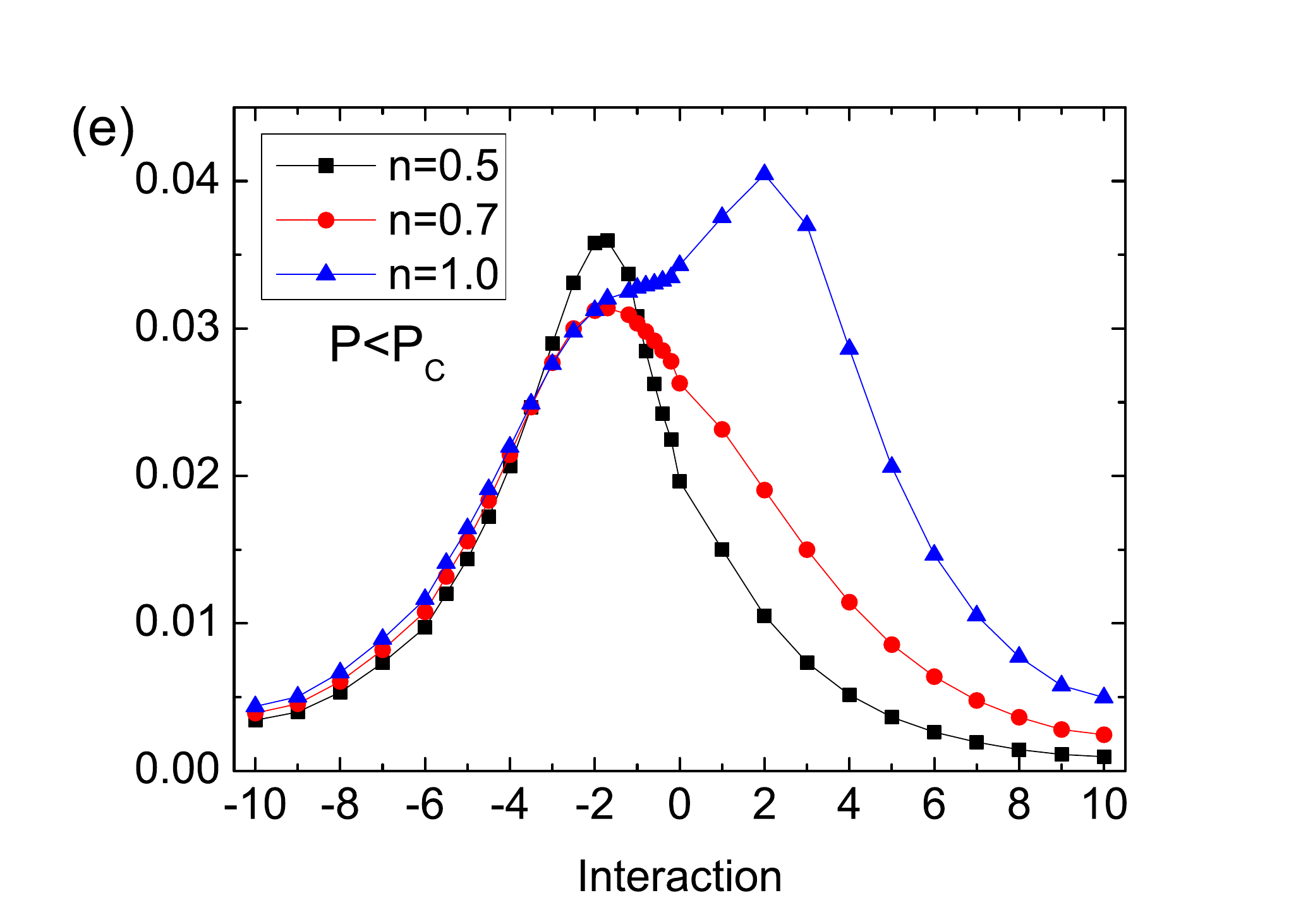}\hspace{-1cm}\includegraphics[scale=0.3]{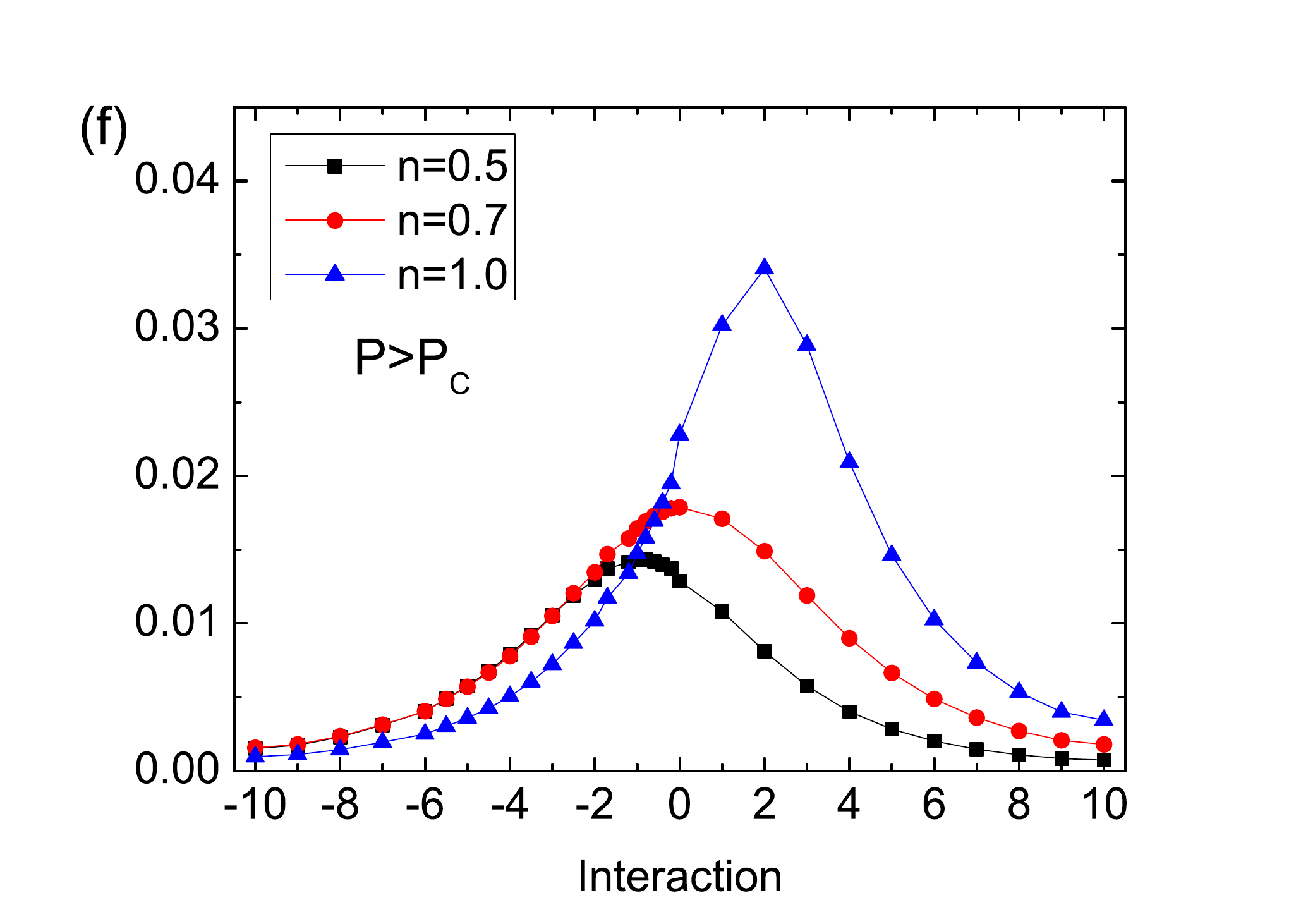}\vspace{-0.5cm}

       \caption{Double-occupancy probability $w_2$ (upper panels) and its derivative $|\partial w_2/\partial U|$ (lower panels) as a function of interaction: (a) and (d) for $P=0$, (b) and (e) for $P<P_C$ ($P=0.1$) and (c) and (f) for $P>P_C$ ($P=0.5$).}
  
     \end{figure*}

\section{\label{sec:level2}Results}

We start by monitoring both magnetic susceptibility and entanglement for a vast regime of parameters: attractive and repulsive interactions, within several filling factors and regimes of polarization, as shown in Figure 1. The most surprising feature is the fact that for moderate and strong attractive interactions ($U\lesssim -2t$) with low polarization ($0\leq P<P_C$) the mapping between $\chi$ and $S$ for any filling factor is the simplest possible: {\it {directly} proportional}. This linear behavior disappears for $P>P_C$ and simply does not occur in any of the repulsive cases. This result suggests that entanglement could be estimated via experimental measurements of the magnetic susceptibility in both conventional ($P=0$) and exotic ($P<P_C$) superfluids.  

To understand this linear mapping between $\chi$ and $S$ we first analyse the entanglement as a function of interaction, in the upper panels of Figure 2. We observe that for any $P$ and $n$ the maximum entanglement occurs at $U=0$. Entanglement thus decreases with $|U|$ monotonically for both attractive and repulsive regimes. This reflects the fact that at $U=0$ the four occupation probabilities are in their best balance for a given filling factor and magnetization, while for $|U|$ increasing there are restrictions in the degrees of freedom via the suppression of the unpaired probabilities in conventional and exotic superfluids, and of the doubly-occupied probability in metals. Note that for $P<P_C$, Figs. 2a and 2b, entanglement is also monotonic with density for the attractive regime and for low repulsive interactions (up to $U\sim3t$ for $P=0$ and up to $U\sim 2.5t$ for $P=0.1$). 

In particular for $U<0$, we find that the impact of the interaction on the entanglement is very similar for all densities. In contrast, for the repulsive regime $U$ increasing leads to a greater decreasing of the entanglement at half filling ($n=1.0$) than at other filling factors, given rise to the non-monotonic behavior of $S$ with $n$. This is a direct consequence of the Mott metal-insulator transition at $n=1$ and $U>0$: there is an energy gap due to the repulsion of electronic charge \cite{shiba}, with a consequent reduction of the degrees of freedom, with maximum entanglement \cite{v1, v15} at $n\sim 0.8$. Interestingly though for $P>P_C$ (Fig.~2c) entanglement is non-monotonic with $n$ for any $U>0$ and also in the attractive regime. The reason is that such strong polarization ($P=0.5$) induces an additional reduction of the degrees of freedom, now related to the spin character through the Pauli exclusion principle. So it can be thought as a spin repulsion effect.

{
Consistently,} the lower panels of Fig.~2 reveal that the non-monotonicities of $S$ with $n$ $-$ due to charge or spin repulsion $-$ {have their} counterpart in the magnetic susceptibility. For repulsive interactions, $\chi$ increases monotonically with $U$ increasing {for both} metallic ($U>0$, $n\neq1$, any $P$) and insulator ($U>0$, $n=1$, any $P$) regimes. This general feature, predicted by Shiba \cite{shiba} for $U>0$ and $P=0$, is here proved to hold also in polarized repulsive systems.

On the other hand, for attractive interactions we find that $\chi$ is non-monotonic with $U$: it has minimum at $U=U_C\sim -2t$ and saturates for $U\rightarrow -\infty$ at finite values, for both conventional (Fig.~2d) and exotic (Fig.~2e) superfluid regimes. This finite saturation in the magnetic susceptibility is actually one of the signatures of the Meissner effect in superconductors \cite{fetter},  due to the coexistence of superconductivity and antiferromagnetic ordering. Accordingly, for the normal non-superfluid regime, i.e. for $P>P_C$ (Fig.~2f), since there is no Meissner effect we observe an almost vanishing $\chi$ for $U\rightarrow -\infty$. 

Compiling all these properties of entanglement and magnetic susceptibility we conclude that the linear mapping between the two quantities for $U<0$ and $P<P_C$ (Fig.~1) is not a coincidence or an artefact. Instead it reflects the similarities between the way both $\chi$ and $S$ respond to $U$ and $n$ (compare Figs. 2a and 2b to Figs. 2d and 2e), while they respond differently to $U$ and $n$ for $U>0$ at any $P$ and for $U<0$ with $P>P_C$. The exception to this is the regime of weakly attractive interactions, $-2t\lesssim U<0$, at which the minimum $\chi$ lies, while the maximum entanglement is at $U=0$, consequently the {linear} relation fails.
   
To investigate the peculiarities of this weakly attractive regime, we first map Shiba's interpretation \cite{shiba} in terms of energy, into the occupation probabilities. According to Shiba, for $U>0$, $\chi$ increases with $U$ increasing and/or $n$ decreasing because the ground-state energy becomes less negative and thus the system can reach magnetization at lower energetic cost. Now in terms of probabilities, increasing repulsive $U$ and/or decreasing $n$ also corresponds to smaller $w_2$, what then implies enhancement of the unpaired probabilities $w_\uparrow, w_\downarrow$. Thus $w_2$ and $\chi$ have opposite behaviors with $U$ and $n$. 
 
The extension of this interpretation to attractive interactions is not straightforward because $w_2$ always benefits from $|U|$ increasing in this case. But the rate at which $w_2$ increases with $|U|$ increasing should be larger for the weak attractive regime (BCS pairs), where $U$ plays a more effective role, than for stronger interactions (strongly coupled pairs, BEC limit), since the impact of $U$ must saturate for $U\rightarrow -\infty$ by reaching the maximum $w_2$. This suggests then that the critical $U_C\sim -2t$ corresponding to the minimum $\chi$ is related to the BCS-BEC crossover.

Our interpretation is confirmed in {Figure 3}: for strong interactions the doubly-occupied probability saturates, $w_2\rightarrow 0$ for $U\rightarrow \infty$, where $\chi\rightarrow \infty$, while $w_2\rightarrow w_2^{max}$ for $U\rightarrow -\infty$, where $\chi$ saturates (at zero for normal state and at finite values for superfluids). For $U<0$ and $P<P_C$ the maximum $|\partial w_2/\partial U|$ occurs at $U=U_C\sim-2t$ ({Figs. 3d and 3e}), precisely the interaction for which $\chi$ is minimum. We see that the maximum  $|\partial w_2/\partial U|$ moves to weaker attractive interactions with $P$, appearing at $U\sim 0$ for $P>P_C$ ({Fig.~3f}). For $U>0$ and $n=1$ another peak is observed at $U\sim 2t$, which reflects the metal-insulator transition. This value of $U$ where the Mott transition occurs is consistent with Shiba's prediction \cite{shiba} and also to the one obtained via fidelity susceptibility in the single-band Hubbard model \cite{ref11}.

{Finally, we explore the universality of the linear relationship between magnetic susceptibility and entanglement in conventional superfluids ($P=0$). By comparing the upper panels of Fig.~1 one finds that while $\chi$ scales similarly for distinct densities, the range of $S$ varies with $n$. This reflects the fact that both minimum (for $U\rightarrow-\infty$) and maximum entanglement (at $U=0$) depend on $n$ (see Fig.~2). Hence to make the entanglement of systems with different densities comparable we have rescaled it by $S\rightarrow S-S_{min}$, where $S_{min}\equiv S(n,m=0,U\rightarrow-\infty)$ is analytically obtained as follows.}

{We first calculate the double occupancy, Eq.(3), from the per-site ground-state energy of attractive systems, which within a particle-hole transformation is given by $e_0(n,m=0,U)=Un/2-e_0(n,|U|)$. As $e_0(n,|U|)$ becomes independent on $U$  for $U\rightarrow -\infty$ \cite{fvc}, we find $w_2=n/2$ and, consequently, vanishing unpaired probabilities $w_\uparrow=w_\downarrow=0$ and $w_0=(1-n/2)$, thus obtaining}
\begin{eqnarray}
{S_{min}(n)}{=-\frac{n}{2}\log_2(n/2)-\left(1-\frac{n}{2}\right)\log_2(1-n/2).}
\end{eqnarray}

{Figure 4 shows that all data of $\chi$ as a function of $S-S_{min}$ lie on top of a single line, revealing thus the universality of the linear relation between entanglement and magnetic susceptibility at $P=0$. Therefore by measuring only the magnetic susceptibility in current superfluid experiments one can quantify entanglement via the universal formula}
\begin{equation}
{S(n,\chi)=S_{min}(n)+\frac{\chi-a}{b},}
\end{equation}
{where $a=0.032$ and $b=-0.029$ are the offset and the slope of the linear fitting of Fig.~4. A similar universal relation for the polarized regime remains to be investigated.}

\begin{figure}[t]
\vspace{-0.5cm}
\hspace{-0.4cm}\includegraphics[scale=0.44]{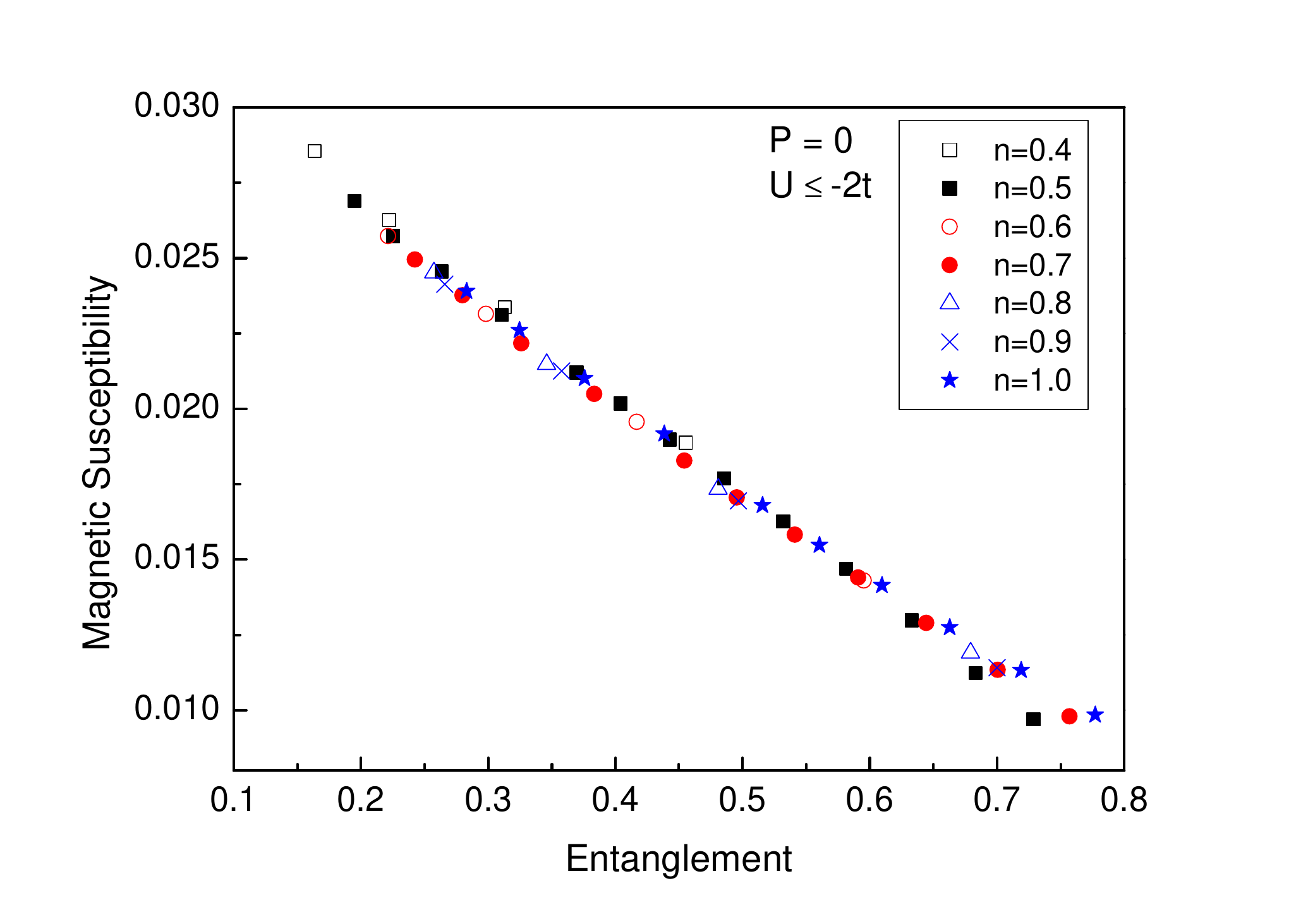}
\vspace{-0.6cm}
\caption{Universal linear relationship between magnetic susceptibility and entanglement in conventional superfluids.}
\vspace{-0.5cm}
\end{figure}

\section{\label{sec:level2}Conclusions}

In summary, we have investigated the mapping between magnetic susceptibility $\chi$ and entanglement $S$ in one-dimensional fermionic systems described by the Hubbard model at zero temperature. We explored a vast regime of interactions $U$, densities $n$ and polarizations $P$, thus comprising the metallic, insulating, conventional superfluid and exotic polarized (FFLO) superfluid phases.  We found a surprising linear mapping between $\chi$ and $S$ in conventional ($P=0$) and exotic ($P<P_C$) superfluids, for $U\lesssim -2t$. We demonstrate that this linearity is neither an artefact nor a coincidence, instead it reflects the similar response of both, entanglement and magnetic susceptibility, to the density and interaction changes. {We have also provided the universal relation between $\chi$ and $S$ in conventional superfluids, thus allowing one to quantify entanglement} by measuring only magnetic susceptibilities in current superfluid experiments. 
     
We found that entanglement is non-monotonic with $n$ for $U<0$ with $P>P_C$ and for $U>0$ with any $P$. While for $U>0$ this behavior is related to the Mott metal-insulator transition, for $U<0$ we attribute this to spin repulsion effects for moderate and strong polarizations. These alike behaviors resemble the similarities between the metallic and the normal non-superfluid phases. 

Finally, our results for the magnetic susceptibility in the attractive interaction regime revealed that for $P<P_C$ $\chi$ is non-monotonic with $U$, with minimum at $U\sim -2t$. By analysing the doubly-occupied probability we showed that this minimum $\chi$ is related to the BCS-BEC crossover. We have also found that $\chi$ saturates with $U\rightarrow-\infty$ at finite values, which is a clear signature of the Meissner effect. In contrast, for the normal non-superfluid phase ($P>P_C$) we found $\chi\rightarrow0$ for $U\rightarrow -\infty$, since the Meissner effect is absent. These very distinct behaviors of $\chi$ for $P<P_C$ and $P>P_C$ could be employed to distinguish between exotic superfluidity (FFLO state) and normal non-superfluid states in spin-imbalanced systems. 

\section*{Acknowledgements}

We thank Guilherme Canella for fruitful discussions. VVF was supported by FAPESP (2019/15560-8) and CNPq INCT-IQ (465469/2014-0). This research was supported by resources supplied by the Center for Scientific Computing (NCC/GridUNESP) from S\~{a}o Paulo State University (UNESP).

\end{document}